\documentclass[review]{elsarticle}
\usepackage{graphicx}
\usepackage{lineno,hyperref}
\usepackage{amssymb}
\modulolinenumbers[5]
\usepackage{color}
\usepackage[table,xcdraw]{xcolor}
\usepackage{multirow}
\usepackage{amsmath,amssymb,amsfonts}
\begin{document}
\begin{frontmatter}
\title{Liver Fat Quantify Network with  Body Shape}
\author[1]{Qiyue Wang\corref{cor1}}%
\ead{wangqiyue@gwu.edu}
\author[2]{Wu Xue}
\author[2]{Xiaoke Zhang}
\author[2]{Fang Jin}
\author[1]{James Hahn}
\cortext[cor1]{Corresponding author}
\address[1]{Department of Computer Science, The George Washington University, USA}
\address[2]{Department of Statistics, The George Washington University, USA}


\begin{abstract}
It is critically important to detect the content of liver fat as it is related to cardiac complications and cardiovascular disease mortality. However, existing methods are either associated with high cost and/or medical complications (e.g., liver biopsy, imaging technology) or only roughly estimate the grades of seatosis. In this paper, we propose a deep neural network to estimate the percentage of liver fat using only body shapes.  The proposed is composed of a flexible baseline network and a lightweight Attention module. The attention module is trained to generate discriminative and diverse features which significant improve the performance. In order to validate the method, we perform extensive tests on the public medical dataset.  The results verify that our proposed method yields state-of-the-art performance with Root mean squared error (RMSE) of 5.26 \% and R-Squared value over 0.8.  It offers an accurate and more accessible assessment of hepatic steatosis. 
\end{abstract}
\begin{keyword}
\texttt{Liver Fat, Body Shape, Attention}

\end{keyword}

\end{frontmatter}

\section{Introduction}
Fatty liver disease which can be categorised by Non-Alcoholic Fatty Liver Disease (NAFLD) and Alcoholic Fatty Liver Disease (AFLD)~\cite{goceri2016} is a common condition caused by the storage of extra fat in the liver~\cite{pickhardt2018}. NAFLD is increasingly common around the world, and affects up to 25 to 30 percent of people in the United States and Europe~\cite{langner2020large,mendoncca2013contrast}. NAFLD is emerging as the most common cause of chronic liver disease, metabolic syndrome, type 2 diabetes and cardiovascular disease~\cite{abenavoli2015non,clark2013cancer}.  The fatty liver disease is often preventable or even reversible in the early stages with lifestyle changes. Therefore, it is of vital importance to monitor and quantify the liver fat and to assess its severity in order to recognize fat-related functional abnormalities in the liver~\cite{guo2020liver}. The classical approach to describe the severity of fatty liver is the degrees of hepatic steatosis which are usually classified into 3 or 4 levels~\cite{sansom2019steatosis}. However, they are failed to tract your liver condition accurately because the grades of hepatic steatosis varies very slow compared with the percentages of liver fat content. For example, the liver fat may decrease or increase several percents in weeks but the grades of hepatic steatosis may not change according to the grades chart~\cite{abenavoli2015non,cervantes1982liver}. As a result, it is practical and necessary to estimate liver fat percentage for routine healthy monitoring.

There are lots of clinical approaches to quantify the liver fat~\cite{hamer2006fatty}. Currently, liver biopsy is considered the gold reference standard for diagnosis and grading of hepatic steatosis~\cite{yan2014automatic}.  A liver biopsy is a test used to diagnose liver conditions. Tissue samples are removed from your liver and conducted  histological analysis under a microscope for signs of damage or disease.  It is a procedure that has risks for complications as well as discomfort for the patients. To avoid the discomfort of invasive methods. The imaging technologies Computerized tomography (CT) scanning and agnetic resonance imaging (MRI) are used to diagnose the liver content. CT is a non-invasive and widely available tool, which provides assessment of hepatic steatosis using Hounsﬁeld Units (HU) associated with each voxel~\cite{hamer2006fatty}.  MRI assesses hepatic steatosis based on its ability to estimate the relative magnitude of the signals arising from fat and from water in each voxel~\cite{o2015ins}. Due to the high cost and/or medical complications associated with these approaches, there is an increasing interest in developing reliable, inexpensive and non-invasive biomarkers to detect and monitor the progression of hepatic steatosis. Abdominal Ultrasound, which uses sound waves to produce pictures to evaluate the size and shape of the liver, as well as blood flow through the liver. Normal liver parenchyma is the same as or slightly more brighter than the adjacent kidney and spleen~\cite{zhang2018liver,charatcharoenwitthaya2007role}.  Thus, the presence of steatosis can be inferred if the liver is too bright and/or if liver structures are blurry or poorly visualised.  Ultrasound scans have the advantage of safety, wide availability and little associated patient discomfort. In addition, the cost of abdominal ultrasound is relatively low compared to CT and MRI. However, it remains relatively insensitive to detection of mild steatosis.  Above methods are time-consuming and requires technical expertise to perform and thus is not routinely available in clinical settings. Beyond these accurate leverages, the blood tests such as liver function tests that can measure fat content of liver. For example, fatty liver disease is diagnosed after blood tests show elevated liver enzymes alanine aminotransferase and aspartate aminotransferase~\cite{rinella2015nonalcoholic}. Although serum markers provide approaches to detect overall hepatic steatosis, they have poor sensitivity and specificity and correlate poorly with the hepatic steatosis~\cite{abenavoli2014serum}.


The body shape related parameters have been verified to be the important predictors of a variety of health biomarkers~\cite{lu20193d,ng2019detailed,wang2020region}, such as body composition and cardiovascular risk. As the boom of computer vision and the explosive growth of optical scan technologies (e.g., using time-of-flight, vision or structured light). It make us easier to acquire 3D body shape data (acquired at home, the gym, or the clinic, using commodity optical body scanning systems including on many smart phones~\cite{lu2018accurate,wang2019novel}). Visceral adipose tissue (VAT) is an important intermediary between body shape and liver fat~\cite{abenavoli2015non,ramirez2018liver} and make it possible to assess the liver fat with body shape.  For example, shape-based anthropomorphic measurements (e.g., body shape index, body roundness index, waist-to-hip ratio and waist to height ratio) are associated with adiposity and can predict the risk of NAFLD~\cite{ramirez2018liver}.   Some previous research have verified the body shape canbe used to predict the degrees of the fatty liver~\cite{wang2021S2FLNet}. However, they only provide the pipeline to classify the grades of the hepatic steatosis and do not accurate assess the percentages of liver fat.
 
Consequently, we propose a shape based \emph{liver fat}percentage prediction \emph{network} which provides a convenient and economical liver fat regression methodology. 

This paper has three main contributions. 

\begin{itemize}
\item We propose a body shape based neural network to quantify the liver fat, which makes the liver fat estimation widely accessible. 
\item We design a novel multi-attention module which learn the discriminative and diverse features via embedding a regression and classification component.
\item We hybrid the attention module along with the baseline regression network to enhance the  representation of the rich attention patterns to improve the regression performance. 
\end{itemize}


\begin{figure*}[h]
    \centering
    \includegraphics[width=\columnwidth]{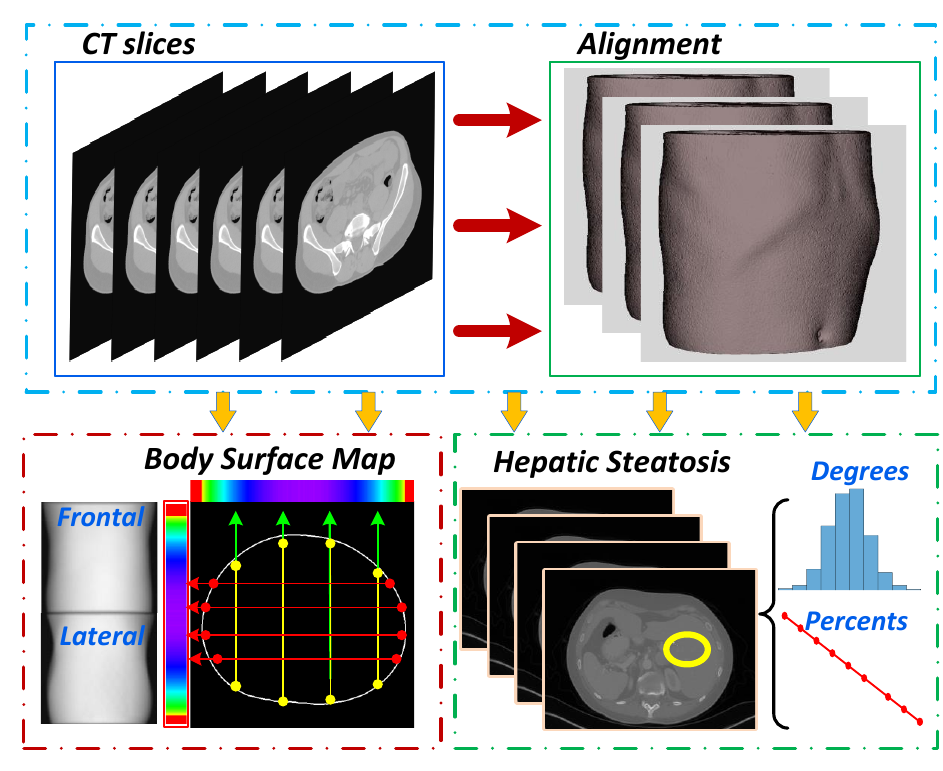}
    \caption{The extraction flowchart of 2D body shape maps and liver fat. The original CTs are calibrated and aligned to the same standard with the pipeline. The frontal and lateral shape maps are extracted from each slice are then combined together. The percentage of liver fat and hepatic steatosis degrees are used as the ground truth is estimated by averaging HU values of different liver regions.}
    \label{fig_C6_1}
\end{figure*}

\section{Related Works}
Body shape related parameters such as The waist circumference, waist-hip ration and BMI have been widely used to assess health biomarkers includes liver fat~\cite{fan2018association}. Because of the poor correlation and simple models, these parameters only used to initial predict the the severity assessment of fatty liver/steatosis. They can not accurate estimate the percentage of the liver fat for continuous monitoring.

Accessibility of 3D human body scanning technologies provide researchers with ways to extract numerous accurate shape related measurements from 3D geometry. In order to investigate correlations between the body shape and various healthy biomarkers~\cite{wang2019novel,ng2019detailed}, Xie et al.~\cite{wang2021pixel} studied body silhouettes and analyzed the correlation between the variation of shapes and the body leanness indicators; Lu et al.~\cite{lu2018accurate} proposed a body fat percentage estimation model by exploring 3D body shape features without anatomical presuppositions. Previous research has shown that VAT is an important intermediary between body shape and liver biomarkers~\cite{abenavoli2015non,ramirez2018liver}. There have been many approaches to predict body fat and VAT volume from the characteristics of body shape~\cite{lu20193d,lu2018accurate,wang2020region,wang2019novel}. Ng et al.~\cite{ng2019detailed} extracted the shape representation from the 3D body surface with principal component analysis (PCA) and provided clinically relevant information to body composition estimates (regional fat/lean masses/VAT). However, these methods adopted simple statistical machine learning models (i.e., Gaussian Process Regression (GPR), Support Vector Machine(SVM), Logistic Regression(LR) etc.) with the discretized shape descriptors. Thus, they are not able to extract deeper shape features resulting in poor predictability. Wang et al.~\cite{wang2021pixel} used dilated residual neural network to extract deep and refined features from 2D body shape maps that works well in predicting the 2D pixel level body composition maps.  In this work, we further explore shape descriptors derived from 3D geometries and propose to use deep learning model to map the body shapes to hepatic steatosis.

\section{Dataset and Preprocessing}

Our aim in this study is to ultilize body shape generated by optical scanner to predict the percentage of liver fat. However, optical body scans with the associated liver fat data are not available at this time. CT scan is an accurate 3D imaging method to provides liver fat estimation and 3D body shapes as well~\cite{abenavoli2015non,pickhardt2018}. The spatial reconstruction precision of CT scan is very high (0.5mm - 2mm) compared to commercial optical body scanners ($\sim$ 2mm). Therefore, we use CT to generate the iso-surface representing the shape of the body as well as calculate the liver fat for the training set in this study. The main contribution of our work is to provide a validated methodology to utilize body shapes to predict the liver fat percentage. The methodology can be used to map body shapes from a variety of sources including optical scanners.

\subsection{Dataset}\label{chpter:6:sec::dataset}
In this study, we use the CT scans from the Cancer Imaging Archive (TCIA)~\cite{clark2013cancer,roth2014new}, and George Washington University Hospital. We manually align all the CT scans and re-sample to ensure the thickness and resolution consistency. We processed all the CT subjects one by one to ensure the reliability of ground truth. Each subject data is reviewed at least twice. The dataset (Fig.~\ref{fig_C6_1}) contains a total of 315 CT scans, which is relatively large in this research field and enough to train the model. 

\subsection{Body Shape Maps }
The reconstructed 3D body shapes by CT iso-surfaces are shown in Fig.~\ref{fig_C6_1} (upper left).  We project the 3D body surface into the 2D body shape from frontal and lateral directions to generate the corresponding 2D body shape maps. The protocol is as follows: we first extract the body contour of each slice and then calculate their depth values. Following this, we combine all the resultant slices to obtain the 2D body shape maps in Fig.~\ref{fig_C6_1} (bottom left). In this study, the image size of the 2D body shape maps is $512 \times 512$.  The detail of the extraction process of body shape maps are included in~\cite{wang2021pixel,wang2021S2FLNet}

\subsection{Liver Fat Results}
Non-contrast-enhanced CT imaging is a non-invasive and widely available modality, which provides objective measurement of x-ray attenuation in Hounsfield Units (HU). The absolute measured attenuation values at each voxel can be used to evaluate the fat content of liver cell ~\cite{pickhardt2018,goceri2016,guo2020liver}. For each subject, we collect at least eight liver regions in different slices as shown in Fig.~\ref{fig_C6_1} (bottom right) and then average their HU values. The liver fat percentages is calculated by the formula which is widely used in~\cite{pickhardt2018}. We also categories all the subjects into four groups with their steatosis grades (grade 0, grade 1, grade 2, grade 3)~\cite{sansom2019steatosis} based on their liver fat percentages.

\begin{figure*}[h]
    \centering
    \includegraphics[width=0.8\columnwidth]{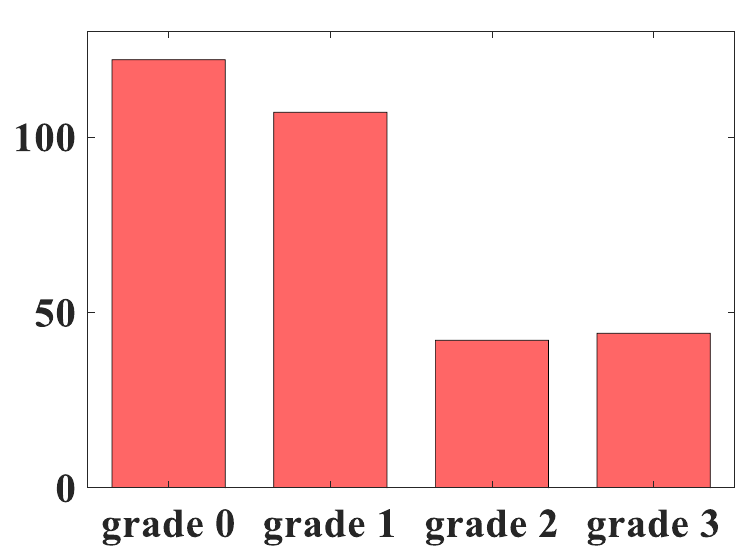}
    \caption[The distribution of steatosis grades.]{The distribution of steatosis grades. The number of each steatosis grades (grade 0, 1, 2, 3) are 122, 107, 42, 44.}
    \label{fig_C6_2}
\end{figure*}

\section{Methodology}
Our objective is to train a deep neural network which can quantify the liver fat percentages with only body shape. The essential of the proposed network is a deep regressor. The inputs and outputs of the model are introduced in Section~\ref{chpter:6:sec::dataset}.

\begin{figure}[t]
    \centering
    \includegraphics[width=1\textwidth]{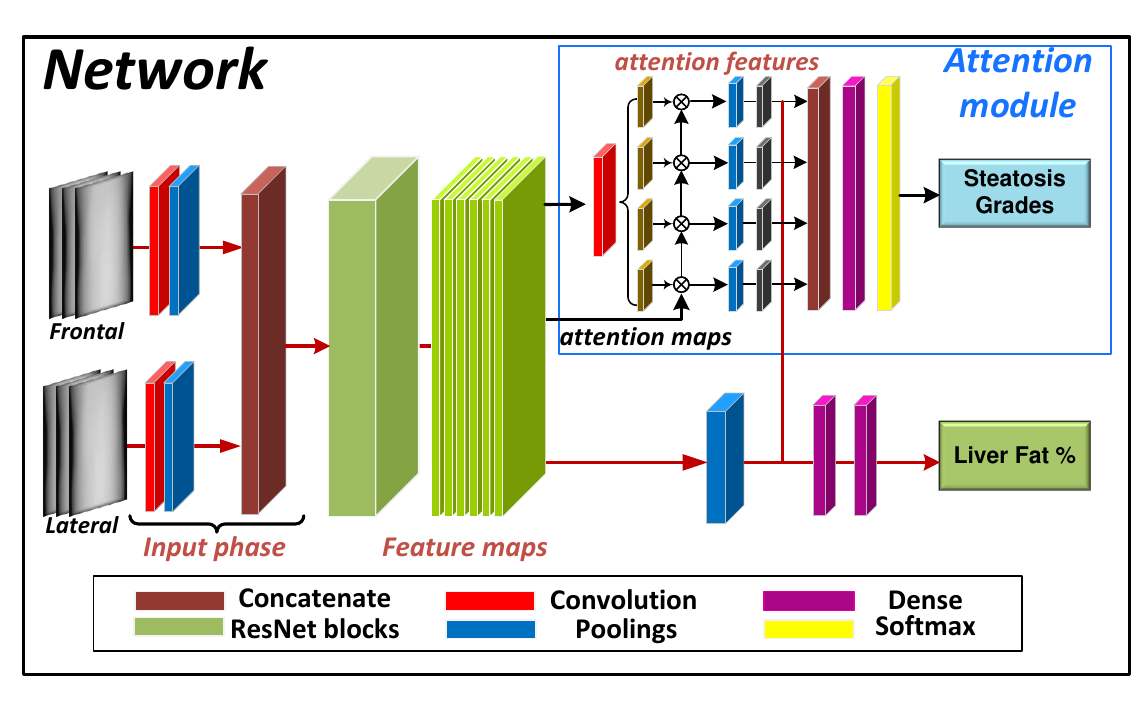}
    \caption{The architecture of the proposed network. The network includes a baseline regression network and a multi-channel attention module. The input of the network is two body shape maps frontal and lateral body shape maps, the outputs of network are liver fat percentage and steatosis grades.}
    \label{fig_C6_3}
\end{figure}

\subsection{Regression Network}\label{chpter:6:sec::regnet}

Fig.~\ref{fig_C6_3} illustrates the overall framework of proposed network for liver fat percentage prediction with body shapes. The proposed network is composed of a baseline backbone network and a multi-channel Attention Module. The baseline neural network contains two inputs: one is frontal body shape map and the other is lateral body shape map. The two input body shape maps are feed into two independent convolution blocks (input phases). Then we concatenate the outputs of the two input phases and then feed it into a classical convolution neural network (CNN) block to generate and extract the deep features. The block is a classical CNN and can be easily transferred from current various networks ~\cite{ioffe2015batch,chen2020deep,roth2015deeporgan}. In this paper, we adopt residual network (ResNet)~\cite{chen2020deep} as the backbone of our model to extract the feature maps .

We make input body shape maps go through the input phase and ResNet blocks to extract the original feature maps denote as $\mathcal{F}_{res}$. The original feature maps are feed into a global average pooling layer (GAP) to generate a one dimension feature vector $V_{GAP}$. Then the feature vector $V_{GAP}$ is used to predict the liver fat percentage via two fully connected layers. This is a typical regression convolution neural network. The whole process is formulated in Eq.~\eqref{eq_C6_1}
\begin{equation}
    \mathcal{P}_{fat}=\Theta(V_{GAP}),\quad V_{GAP}=GAP(\mathcal{F}_{res})
    \label{eq_C6_1}
\end{equation}
where the $\mathcal{P}_{fat}$ is the liver fat percentage and the $\Theta(\cdot)$ is the regression process operation and the $GAP(\cdot)$ is the GAP operation.

\subsection{Attention Module}

\begin{figure}[t]
    \centering
    \includegraphics[width=1\columnwidth]{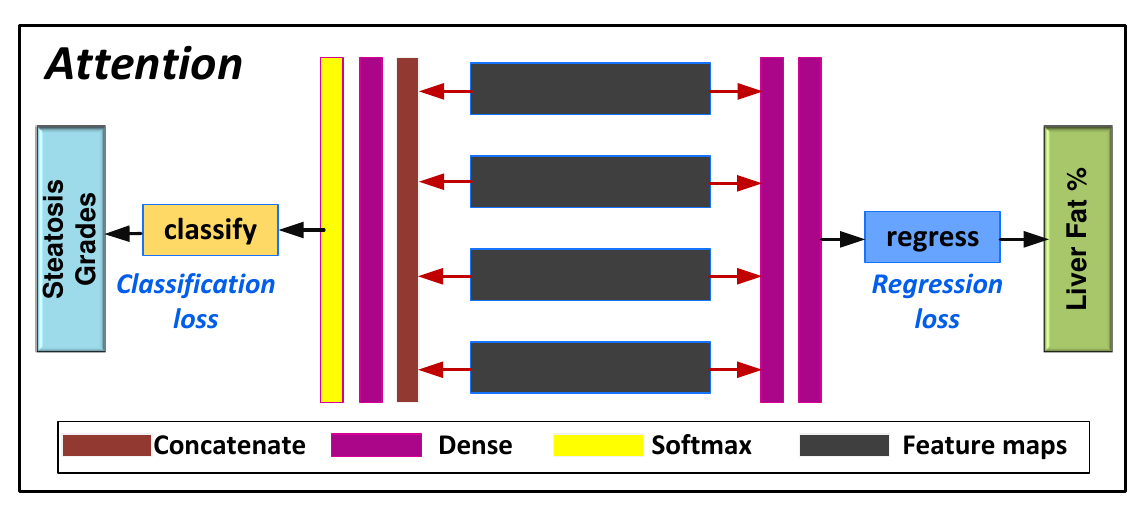}
    \caption{The detail structure of attention module. The attention module owns four channel whihc is consistent with the steatosis grades. It consists of two component: classification and regression component.}
    \label{fig_C6_4}
\end{figure}

The input of our network is two body shape maps with dimension $512\times512$, which is a relative large image size. Not all regions or pixels of the body shape maps are significant for liver fat prediction. We want to focus more on effective region which contribute more on predicting liver fat percentage. Although the baseline regression network owns ability to extract features, but as the limitation of model depth, it is not effective to generate discriminative and diverse feature maps.
The attention mechanism is one of the most valuable breakthroughs in deep Learning research in the last decade. Attention module is a technique that mimics cognitive attention, is also used to make deep neural network learn and focus more on the important information, rather than learning non-useful background information. 

We design and introduce a attention module to improve the performance of the proposed network. The attention module is expected to learn more information on distinguish steatosis grades. Thus the channel of the attention is same as the steatosis grades (grade 0$ \sim $3), as introduced in Section~\ref{chpter:6:sec::dataset}.
The multi-attention module is trained to refine the feature maps, thus we insert it after the ResNet blocks where we get the deep feature maps by ResNet. The input of the attention module is the feature maps $\mathcal{F}_{res}$. Taking the feature map $\mathcal{F}_{res}$,  we generate four attention maps to indicate the distributions of attention on four different steatosis grades. After that, we element-wise multiply input deep feature maps $\mathcal{F}_{res}$ by each attention map to generate four refined feature maps denoted as ${F}_{att}$, as shown in Eq.~\eqref{eq_C6_2}.

\begin{equation}
    \mathcal{M}_{att}^k=\Psi(\mathcal{F}_{Res}),\quad 
    \mathcal{F}_{att}^k={M}_{att}^k\cdot\mathcal{F}_{res},\quad  k=1,2,3,4
    \label{eq_C6_2}
\end{equation}
where $\mathcal{M}^k_{att}$ is the $kth$ attention map generated by$\Psi()$, $\mathcal{F}_{att}^k$ is the $k_{th}$ refined feature maps generated by attention module.

The four attention feature maps $\mathcal{F}_{att}$ are feed into two different components to improve the diversity and discriminability of proposed network. Fig.~\ref{fig_C6_3} shows the details architecture of our attention module. Our attention module is separated into two components one is classification component another is regression component. The classification component is to distinguish the steatosis grades and enhance the diversity of the model. The regression component is to predict the liver fat percentages and improve discriminability correspondingly. The regression component and baseline regression network share the final two fully connected layers.

The attention module is trained together with our baseline regression network. As the Fig~\ref{fig_C6_4}, the feature map $\mathcal{F}_{res}$ of the regression network which is used to predict the liver fat is trained to learn the rich attention representation of body shapes via attention mechanism. Thanks to the parameter-sharing and backward-propagation, the last two fully connected layers is trained with attention module. Therefore, after the model is trained, we do not use the attention module and can only use the backbone of the regression network for liver fat prediction. Consequently, there is no additional computation cost at prediction phase which makes our proposed network more efficient solution for practical usage.

We adopts a structure of the form BatchNorm-ReLu~\cite{ioffe2015batch} in to our network. It can accelerates the training process and has the characteristic of regularization to reduce over-fitting.

\subsection{Loss Functions}
 The proposed network is a form of multi-task network and is simultaneously optimized by baseline regression network and attention module. Specifically, we construct the loss function as a multi-task optimization problem, which is defined as follows:
  \begin{equation}
    \mathcal{L}_{total}= \lambda_1\mathcal{L}_{reg}+ \lambda_2\mathcal{L}_{att}
    \label{eq_C6_3}
\end{equation}
where $\mathcal{L}_{total}$ is the joint loss of our network. $\mathcal{L}_{reg}$ and $\mathcal{L}_{att}$ are the loss of baseline regression network and attention module and $\lambda_1, \lambda_2$ are the loss weights of them correspondingly. The regression loss in our model is mean squared error (MSE).

As shown in Fig.~\ref{fig_C6_4} and Section \ref{chpter:6:sec::regnet}, our multi-attention module is composed by two different components: regression component, classification component. The regression component and the classification component have different loss functions. The loss of regression component is also a MSE loss which is similar with the $\mathcal{L}_{reg}$ in baseline regression network. The classification component is a typical classification model, thus we select its loss to a cross entropy. The Eq.~\eqref{eq_C6_3} is rewritten as:
  \begin{equation}
    \mathcal{L}_{total}= \lambda_1\mathcal{L}_{reg}+ \alpha_1\mathcal{L}_{att-reg}+
    \alpha_2\mathcal{L}_{att-cls}
    \label{eq_C6_4}
\end{equation}
where $\mathcal{L}_{att-reg}$ and $\mathcal{L}_{att-cls}$ are the loss of regression component and classification component respectively. $\alpha_1,\alpha_2$ are the loss weights of $\mathcal{L}_{att-reg}$ and $\mathcal{L}_{att-cls}$ correspondingly. When training the model, we minimize the joint loss $\mathcal{L}_{total}$ and get the target outputs. 

\section{Experiment}

In this section, we evaluate the performance of the liver fat quantification of our proposed deep neural network and validate it on the medical datasets.

\begin{figure}[t]
    \centering
    \includegraphics[width=1\columnwidth]{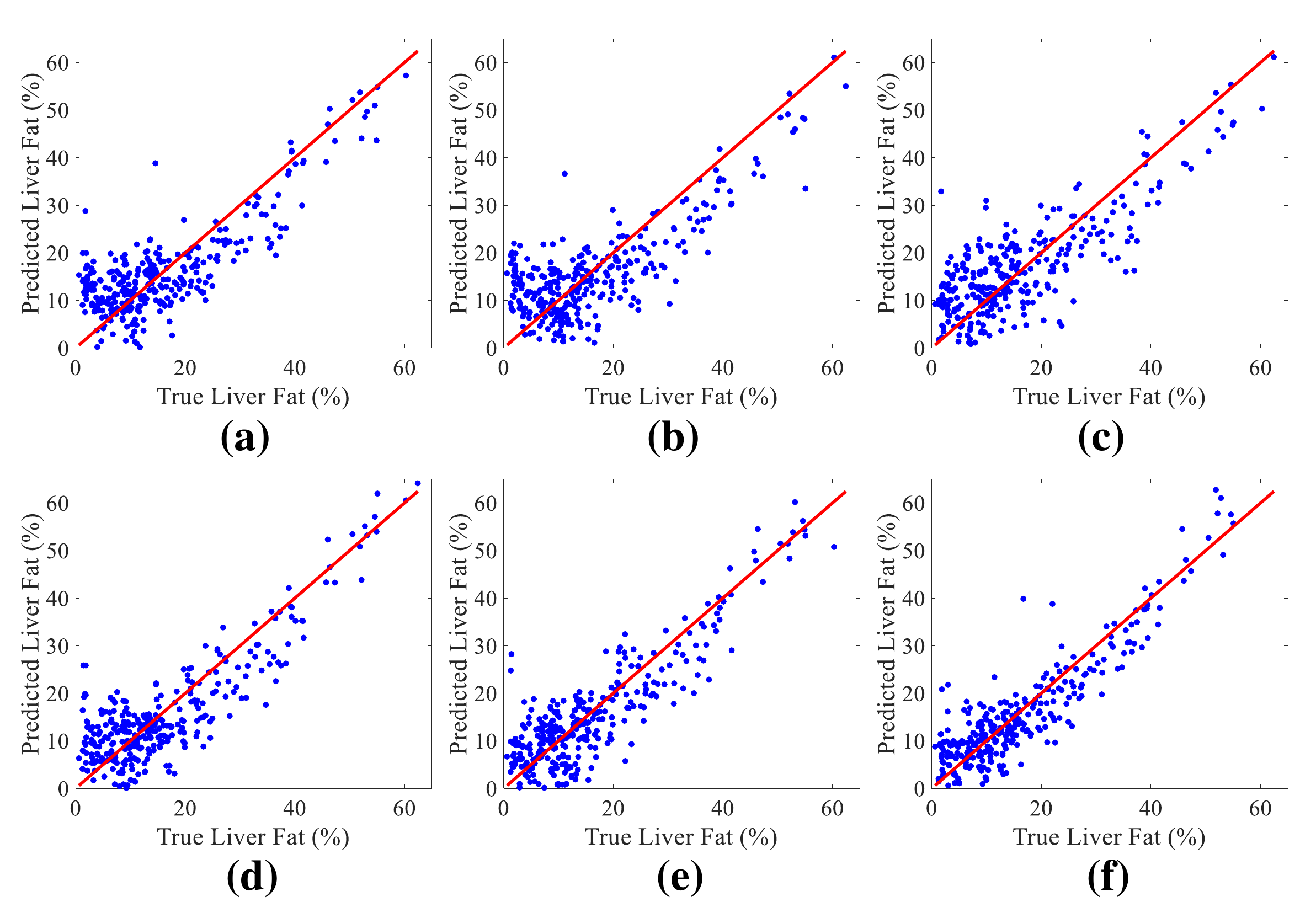}
    \caption{The predicted and truth liver fat percentages results of different methods. From (a) - (f) are results of Linear Regresssion, SVM, MLP, ResNet, Baseline and proposed method.}
    \label{fig_C6_5}
\end{figure}

\subsection{Implementation}
The proposed network and all reference methods in this paper are implemented on the open source deep learning framework tensorflow. To optimize the our network, we select the stochastic gradient descent (SGD) optimizer with initial learning rate of 0.01. We set the learning rate decay by 0.1 after every 20 epochs and also set the momentum equals to 0.9.  We set the mini-batch size to 32 and the overall training phase to be 200 epochs. The training and validation processes are performed on two NVIDIA GTX 1080Ti graphics cards (11GB GPU memory for each). 

To evaluate the performance of each method comprehensively, we calculated the root mean square error (RMSE) and R-Squared to evaluate the regression performance. We also use classification accuracy,  and confusion matrix to validate the steatosis grades based on regression results. Due to the data size limitation of medical dataset, and to reduce the impact of particular random choice of samples, we use five-fold cross validation to evaluate the performance of the methods.

\renewcommand\arraystretch{1.5}
\begin{table}[t]
\centering
\caption{Comparison results with the reference methods.}
\begin{tabular}{c|c|c|c}
\hline
{\bfseries Methods}  & {RMSE\hspace{0.2cm} }& {R-Squared} & {Steatosis Grades Accuracy $\%$}\\
\hline
Linear Regression                     & 7.04     & 0.552 	 & 53.3 	 \\
SVM                                   & 7.61     & 0.453 	 & 50.8 	 \\
MLP~\cite{manaswi2018regression}      & 7.24     & 0.542 	 & 53.7 	 \\
ResNet~\cite{chen2020deep}            & 6.67	 & 0.663      & 57.5  	 \\
Baseline Network                     &  6.13	 &  0.739     & 63.2 \\
\hline
\bfseries Proposed Network          & \bfseries 5.26	   & \bfseries 0.815  & \bfseries  72.1 \\
\hline
\end{tabular}
\label{tab_C6_1}
\end{table}

\subsection{Performance Comparison with Reference Methods}
We compare to several methods to validate the effectiveness of our proposed network. The majority of previous of body shape based biomarkers assessment adopted statistic models~\cite{lu20193d,wang2019novel,ng2019detailed}. Therefore, we apply the Linear Regression and the Support-Vector Machines (SVM) which are popular in this field. Our inputs are two $512\times512$ images, which are raw features with high dimension.  Thus, they are not appropriate for direct use in Liner Regression and SVM. Therefore, we apply PCA on our 2D body shape maps and select the top principal components (PCs) of our body shape maps which characterize at least $95\%$ of the variation in body shape. Then model the Linear regression and SVM with selected top PCs.  In order to validate our method, we also compared our method with other popular state-of-the-art neural network methods, multi-layer perceptron (MLP), ResNet, baseline method. The baseline method shares the merely same architecture with our proposed network but only include the regression attention component. Therefore, we have at least two specific aims: one is to verify the effectiveness of our multi-channel attention module another is validate the superiority deep neural network.

\begin{figure}[t]
    \centering
    \includegraphics[width=1\columnwidth]{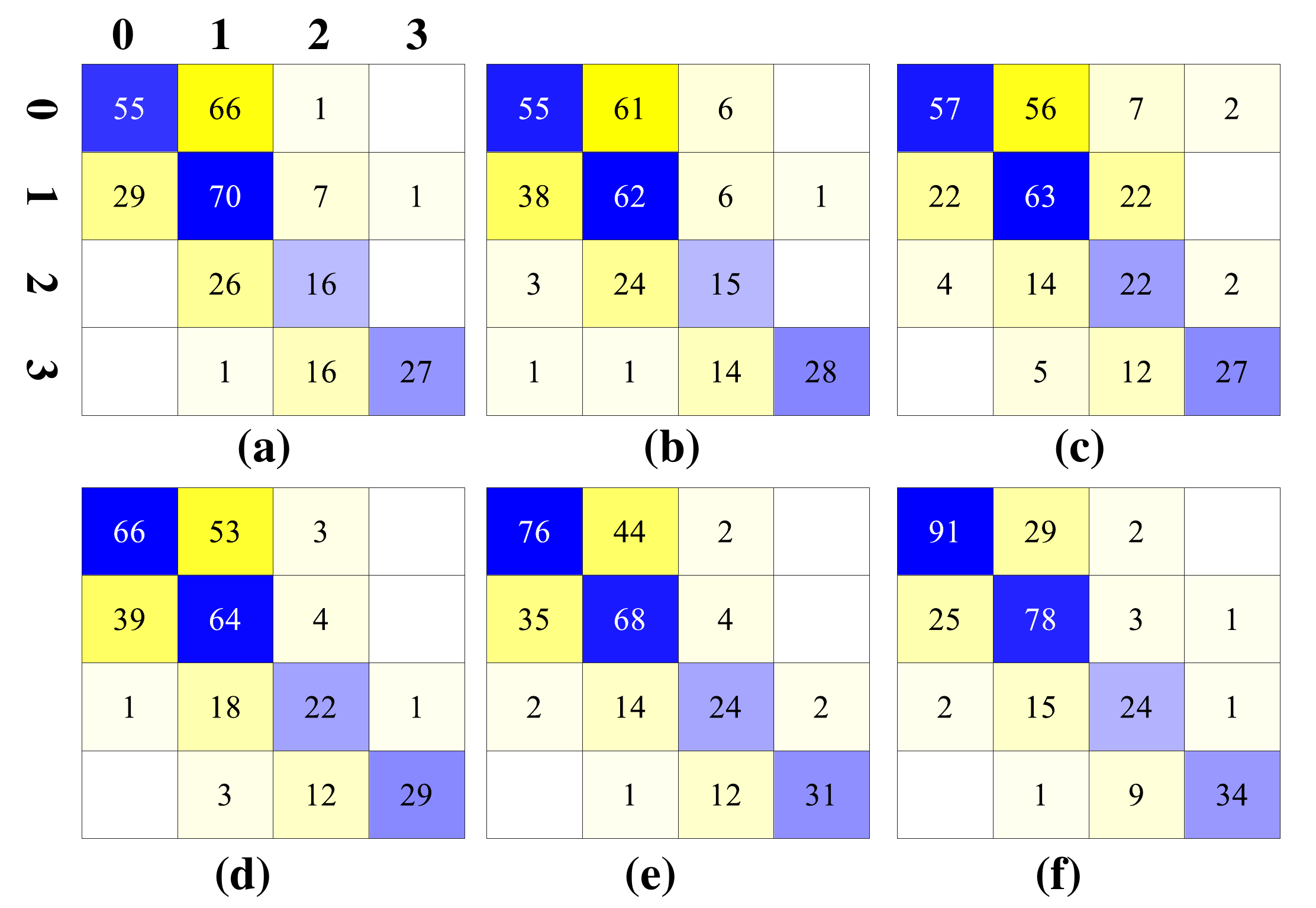}
    \caption{The confusion matrix results of different methods. From (a) - (f) are results of Linear Regresssion, SVM, MLP, ResNet, Baseline and proposed method.}
    \label{fig_C6_6}
\end{figure}

We run all the methods on the medical dataset and calculate all the metrics which show at the Table~\ref{tab_C6_1}. Table~\ref{tab_C6_1} lists the RMSE, R-Squared and grading accuracy of steatosis degrees. We can observe that deep neural network surpass the classical models in terms of these three metrics. Compared with Linear Regression, SVM and MLP, the baseline and our proposed network significant improve the performance.  As mentioned above, both baseline and proposed network include a attention mechanism. Their performance obviously better than other reference methods. It is worth noting that proposed network which embedded our multi-channel attention module work best and its R-squared is 0.815 and over 0.8. It means the proposed network can estimate the liver fat percentage very well. Compared with baseline which does not includes classification component in attention, our proposed network improve the performance about $15\%$.

We draw the predicted $vs$ truth diagram shows in Fig.~\ref{fig_C6_5} to further demonstrate the performance. The Fig.~\ref{fig_C6_5} is consistent with the results in Table~\ref{tab_C6_1}, and our proposed network estimate the liver fat very well. All the methods especially the statistical models tend (Fig.~\ref{fig_C6_5}-a, Fig.~\ref{fig_C6_5}-b) to over estimate the liver fat in low steatosis grades which are typically in liver fat estimation~\cite{goceri2016}. Our baseline and proposed method has a slighter over estimation compared with them. We can also observe this in Fig~\ref{fig_C6_6}. 
We also compared the steatosis grades classification based on the results of liver fat percentages to show the robust of the models. The total classification accuracy is shown in last column of Table~\ref{tab_C6_1}. The corresponding confusion matrix is shown in Fig.~\ref{fig_C6_6}. The classification performance is similar with the regression results. Our proposed network surpasses all other methods in total accuracy and the accuracy of each grade. The main strengths of our proposed method is the improving in low steatosis grades which we also observed in Fig.~\ref{fig_C6_5}. 

These experimental results demonstrate that the 2D body shape maps canbe sued to predict the liver fat. The deep neural networks surpass the classical statistical models because their the deep feature extraction ability. In addition, according to the results of different network variations, the attention mechanism in baseline network and proposed network can improve the performance. Our proposed multi-channel attention modules can significantly enhance the regression and corresponding classification performance.

\subsection{Visualization and Interpret}
\begin{figure}[t]
    \center
    \includegraphics[width=\columnwidth]{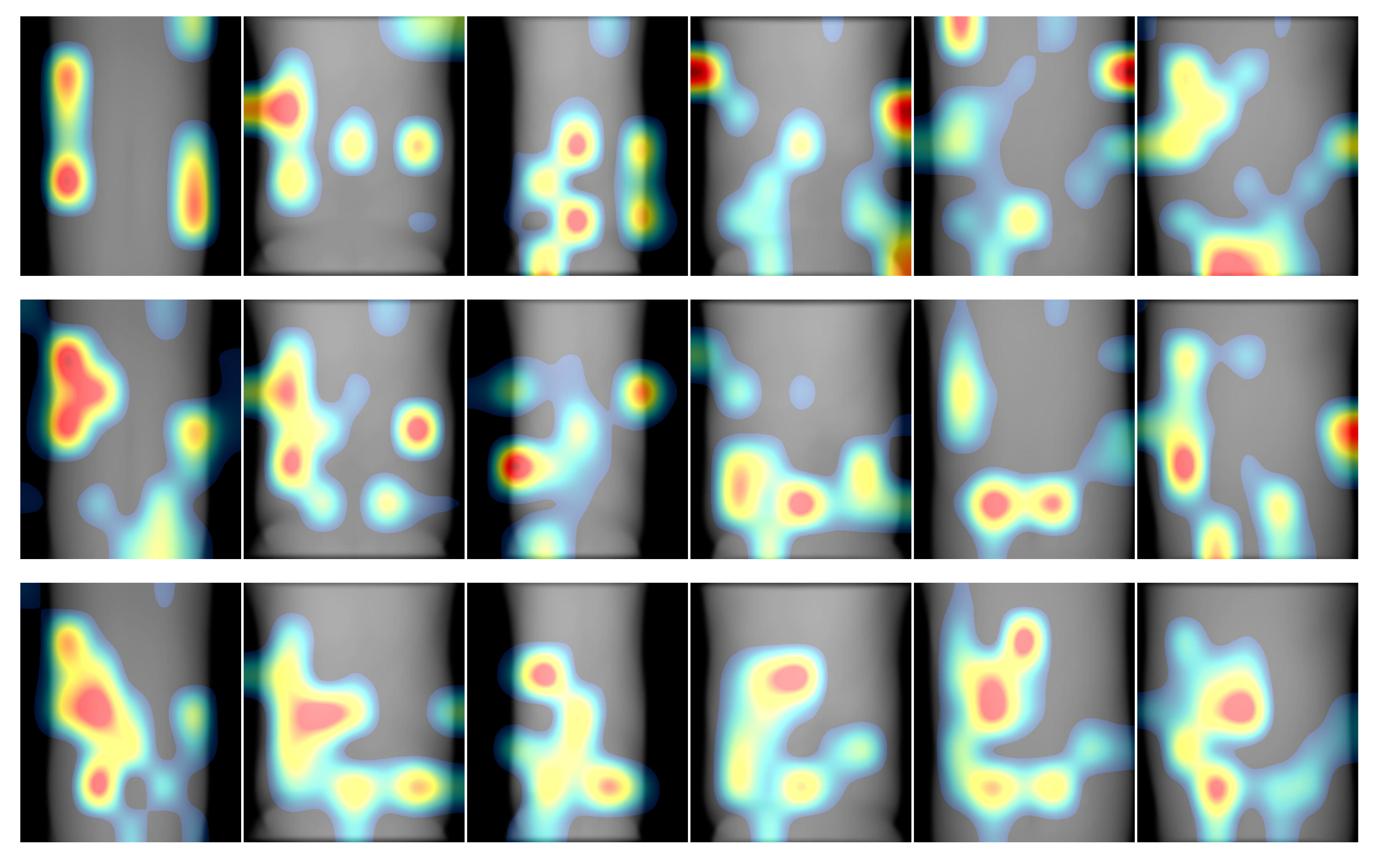}
    \caption [Heatmap visualization]{Heatmap visualization. The Top row shows the heatmaps of the ResNet, the mid row shows the heatmaps of baseline method and the bottom row shows the heatmaps of the proposed network overlaid on top of the frontal map of the body. The highlighted regions refer to the effective regions to quantify the liver fat, the warmer color means higher correlation.
    }
    \label{fig_C6_7}
\end{figure}

The one major weakness of deep neural network is the lack of interpretable. To make our deep neural networks more interpretable. We conduct visualization experiments to investigate the effective body regions used by the deep neural network to quantify the liver fat. This is designed to give insight on the inner workings of the proposed network that would allow clinicians to interpret the ``black box" often associated with machine learning approaches. There are some previous works to explore inner explanation of deep convolution neural networks (CNNs)~\cite{zhou2016learning,selvaraju2017grad}. Gradient-weighted Class Activation Mapping (Grad-CAM) ~\cite{selvaraju2017grad} is a simple technique to get the discriminative image regions used by a CNN to identify a specific class in the image. It could also be used to interpret the prediction decision made by the deep neural network.  We generate the feature activation heatmaps by Grad-CAM for ResNet, baseline network and the proposed network. The results are shown in Fig.~\ref{fig_C6_7}. According to the experiment results, the ResNet and the baseline network tends to exhibit a circumscribed activation on the sides of the abdomen which is the profile of the body shape. These regions are good representation of body shapes but not the best to predict liver fat~\cite{wang2021S2FLNet,childs2016ultrasound}. In addition to this, the proposed network tends to activate the waist region and the region where the liver is located. This indicates that our proposed network has the ability to learn from a rich representation of the body shape.

\section{Conclusion}
In this study, we present a novel deep neural network based approach to determining the liver fat percentage using body shapes. In order to improve the prediction performance, we introduce a novel multi-attention module to enhance the diversity and discriminability of the feature maps. State-of-the-art methods and proposed network are tested on the medical datasets.  The experimental results show that the proposed deep neural network is superior to reference methods. The proposed network with designed attention module works best in liver fat prediction and as well as assessment of steatosis grades. The experiment indicates that our proposed network provides a less-expensive and accurate alternative to competing solutions for routine monitor of liver fat.

\section*{Acknowledgements}
This work was supported in part by the National Institutes of Health under grants R01DK129809.
\bibliographystyle{splncs04}
\bibliography{mybibliography}

\end{document}